\documentclass{aa}  

\usepackage{soul}
\usepackage{graphicx}
\usepackage{multirow}
\usepackage{txfonts}
\usepackage{amssymb,amsmath}
\usepackage{xcolor}
\usepackage{stfloats}
\usepackage{placeins}
\usepackage{hyperref}
\hypersetup{
    colorlinks=true,
    linkcolor=blue,
    citecolor=blue,
    urlcolor=blue}
\usepackage[titletoc,toc,page]{appendix}

\begin{document}

   \title{A machine learning approach to estimate mid-infrared fluxes from WISE data}
   %\subtitle{}

   \author{Nuria~Fonseca-Bonilla\inst{1}
          \and
          Luis~Cerd\'an\inst{2}
          \and
          Alberto~Noriega-Crespo\inst{3}
          \and
          Amaya~Moro-Mart\'in\inst{3}
          }

   \institute{Centro de Astrobiolog\'ia (CSIC-INTA), Instituto Nacional de T\'ecnica Aeroespacial, 28850 Torrej\'on de Ardoz, Madrid, Spain\\
              \email{fonsecabn@cab.inta-csic.es}
         \and
             Instituto de Qu\'imica-F\'isica ``Blas Cabrera'', Consejo Superior de Investigaciones Cient\'ificas, 28006, Madrid, Spain%\\
             %\email{l.cerdan@csic.es}
        \and
            Space Telescope Science Institute, 3700 San Martin Drive, Baltimore, MD 21218, USA
        }     

   \date{Received 6 April 2024 / Accepted 25 September 2024}

% \abstract{}{}{}{}{} 
% 5 {} token are mandatory
 
  \abstract
   {While the Wide-field Infrared Survey Explorer (WISE) is the largest, best quality infrared all-sky survey to date, a smaller coverage mission, \textit{Spitzer}, was designed to have better sensitivity and spatial resolution at similar wavelengths. 
   Confusion and contamination in WISE data result in discrepancies between them.}
   {We aim to present a novel approach to work with WISE measurements with the goal of maintaining both its high coverage and vast amount of data while, at the same time, taking full advantage of the higher sensitivity and spatial resolution of \textit{Spitzer}.}
   {We have applied machine learning (ML) techniques to a complete WISE data sample of open cluster members, using a training set of paired data from high-quality \textit{Spitzer} Enhanced Imaging Products (SEIP), MIPS and IRAC, and allWISE catalogs, W1 (3.4 $\mu$m) to W4 (22 $\mu$m) bands. We have tested several ML regression models with the aim of predicting mid-infrared fluxes at MIPS1 (24 $\mu$m) and IRAC4 (8 $\mu$m) bands from WISE variables (fluxes and quality flags). In addition, to improve the prediction quality, we have implemented feature selection techniques to remove irrelevant WISE variables.}
   {We have notably enhanced WISE detection capabilities, mostly for the targets with the lowest magnitudes, which previously showed the largest discrepancies with \textit{Spitzer}. In our particular case, extremely randomized trees was found to be the best algorithm to predict mid-infrared fluxes from WISE variables, attaining coefficients of determination $R^2\sim0.94$ and $R^2\sim0.98$ for 24 $\mu$m (MIPS1) and 8 $\mu$m (IRAC4), respectively. 
   We have tested our results in members of IC 348 and compared their observed fluxes with the predicted ones in their spectral energy distributions. We show discrepancies in the measurements of \textit{Spitzer} and WISE and demonstrate the good concordance of our predicted mid-infared fluxes with the real ones.}
   {Machine learning is a fast and powerful tool that can be used to find hidden relationships between datasets, as the ones we have shown to exist between WISE and \textit{Spitzer} fluxes. We believe this approach could be employed for other samples from the allWISE catalog with SEIP positional counterparts, and in other astrophysical studies in which analogous discrepancies might arise when using datasets from different instruments.}


   \maketitle

%
%--------------------------Intro------------------------------

\section{\label{sec:introduction}Introduction}

Infrared observations have been used to study a wide range of phenomena, more prominently the formation and evolution of planetary systems, exoplanet characterization, the birth and death of stars, interstellar gas and dust, and the formation and evolution of galaxies, including those that formed in the very early universe \citep{GlassBook1999}. 
Many of these observations have been carried out from space using both all-sky surveys and pointed observations. 
The first infrared space mission, the Infrared Astronomical Satellite (IRAS) \citep{Neugebauer1984}, carried out an all-sky survey at 12, 25, 60,  and 100 $\mu$m with a spatial angular resolution of  30$\arcsec$ at 12 $\mu$m to 2$\arcmin$ at 100 $\mu$m.  This selection of wavelength ranges allowed measuring the spectral energy distribution (SED) of a wide range of astrophysical objects, from stars to starburst galaxies \citep{1988ApL&C..27...67B}. 
Since then, other infrared space missions have followed:  ISO \citep{Kessler1996}, \textit{Spitzer} \citep{Werner2004}, AKARI \citep{Murakami2007}, the Wide-field Infrared Survey Explorer (WISE) \citep{Wright2010}, and the \textit{James Webb} Space Telescope (JWST) \citep{Gardner2006}. These missions have aimed to have some overlap in wavelength range that allows them to bootstrap their absolute flux calibration by  observing the same objects at a similar wavelength range to be able to deal with, understand, and mitigate instrumental effects. Nevertheless, by design, the resolving power of their optics (i.e., spatial angular resolution) and sensitivity have not been the same. In this study we want to compare data from two of these infrared space missions:  \textit{Spitzer} and WISE.

\begin{figure*}[tbhp]
\centering
\includegraphics[width=0.8\linewidth]{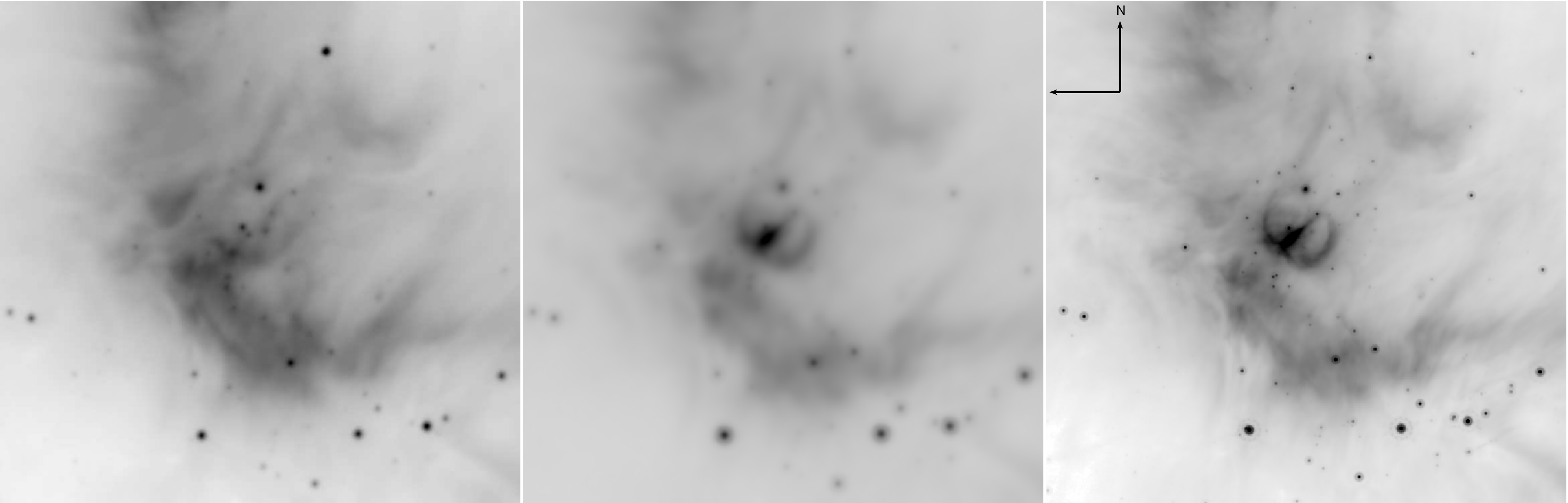}
\caption{Comparison of the open cluster IC 348 observed with WISE W3 at 12 $\mu$m (left), W4 at 22 $\mu$m (center), and \textit{Spitzer} MIPS1 at 24 $\mu$m (right), on a 20 arcmin field of view. Notice that the morphology of the extended emission is different between the 12 $\mu$m (W3) and 22 $\mu$m (W4) images. On the other hand, the 22 $\mu$m and 24 $\mu$m  (MIPS1) data look quite similar overall, except that the difference in sensitivity and angular resolution makes the 24 $\mu$m image look sharper and makes it possible to detect fainter point sources.}
\label{fig:IC348}
\end{figure*}

WISE stands as the last all-sky mid-infrared mission to date. It surveyed the sky at four different wavelengths at 3.4 $\mu$m (W1), 4.6 $\mu$m (W2), 12 $\mu$m (W3), and 22 $\mu$m (W4), with a better sensitivity than any other previous infrared mission that mapped the whole sky photometrically, detecting almost 750 million objects on its images. WISE bands are very similar to some of those of \textit{Spitzer} (Table~\ref{table:instruments}) that made both photometric and spectral observations at different wavelengths between 3 and 180 $\mu$m. 
Nevertheless, \textit{Spitzer} has a better performance concerning resolution, sensitivity, and saturation limit \citep{Cutri2013, Antonucci2014}, see Fig.~\ref{fig:IC348}. 
The \textit{Spitzer}'s 85 cm primary mirror, compared to WISE's 40 cm one, provided \textit{Spitzer} with a resolution twice as good.
WISE achieved 5$\sigma$ point source sensitivities better than 0.08, 0.11, 1, and 6 mJy at 3.4, 4.6, 12, and 22 $\mu$m, respectively, in unconfused regions on the ecliptic in its four bands, with a slightly better sensitivity toward the ecliptic poles \citep{Wright2010}. For comparison, \textit{Spitzer's} IRAC instrument had a point-source sensitivity requirement (5$\sigma$, 200 sec) of 6, 7, 36, and 54 $\mu$Jy at 3.6, 4.5, 5.8, and 8.0 $\mu$m, respectively \citep{Werner2004}. At 24 $\mu$m (MIPS1), the \textit{Spitzer MIPSGAL} survey on a 278 deg$^2$ of the Galactic plane reached a 5$\sigma$ RMS point source sensitivity of 1.3 mJy \citep{Carey2009}; similarly, the \textit{Cores-2-Disks (C2D)} survey on a 10.6 deg$^2$ in Perseus reached a 5$\sigma$ RMS point source sensitivity of $\sim$ 1 mJy \citep{Rebull2007}. Over its mission lifetime, \textit{Spitzer} observed just a small fraction of the sky, compared with the all-sky coverage of WISE.

\begin{table*}
    \caption{\textit{Spitzer} and WISE instrumental characteristics extracted from \citet{Antonucci2014} and \citet{vanderMarel2016}.}            
    \label{table:instruments}      
    \centering                          
    \begin{tabular}{l|cccccc}
    \hline            
    \hline
    \noalign{\smallskip}
    \multirow{2}{*}{Mission} & \multirow{2}{*}{Coverage} & Primary Mirror &
    \multirow{2}{*}{Bands} &  Wavelength & Resolution & Beam Size  \\
            &  & Diameter &  & ($\mu$m) & (arcsec pixel$^{-1}$) & (arcsec) \\
    \noalign{\smallskip}
    \hline
    \hline
    \noalign{\smallskip}
    \multirow{2}{*}{WISE} & \multirow{2}{*}{all-sky} & \multirow{2}{*}{40 cm} & W1, W2, W3 & 3.4, 4.6, 12.0 & 2.7 & 6.1, 6.4, 6.5 \\
        & & & W4 & 22.0 & 5.5 & 12.0 \\
    \multirow{2}{*}{\textit{Spitzer}} & \multirow{2}{*}{targeted} & \multirow{2}{*}{85 cm} & IRAC1-4 & 3.6, 4.5, 5.8, 8.0 & 1.2 & 1.7-1.9 \\
       & & & MIPS1 & 24.0 & 2.6 & 6.0 \\
    \noalign{\smallskip}
    \hline                                   
    \end{tabular}
\end{table*}

This difference in instrumental performance is reflected in that several studies have found indications of some kind of contamination in WISE data \citep{Debes2011,Patel2017,Xu2020}. A number of them have even shown unexpected discrepancies between results in WISE and \textit{Spitzer} \citep{Kennedy2012,Silverberg2018,Dennihy2020}. 
One possible explanation is that the wide point-spread function (PSF) in WISE leads to source confusion, allowing background or nearby emission to contribute to the measured photometry, which contaminates the sample \citep{Patel2017,Silverberg2018,Dennihy2020}. 
Due to the similar band passes of WISE and \textit{Spitzer}, it would be tempting to use \textit{Spitzer} instead of WISE when available, but the ending of the \textit{Spitzer} mission some years ago restricts that solution to just a small area of the sky. 
Thus, an alternative solution is needed to take full advantage of WISE's all sky coverage.

In the last years, driven by the extraordinary advances in technology, the amount of data gathered in astronomical observations has grown exponentially \citep{Ball2010, Pesenson2010, KremerJ2017, Baron2019}. 
This increase in quantity, but also in complexity, has made almost impossible to handle analysis with traditional methods.
Machine learning (ML), the subfield of artificial intelligence focused on the use of data and mathematical algorithms that mimic the way humans learn, offers a variety of different approaches to solve these problems \citep{BishopBook2007, LindholmBook22}.
In fact, the use of ML techniques in astronomy has been so prolific that it has even led to its own research branch \citep{KremerJ2017,astroMLBook, FlukeCJ2020, Moriwaki2023}. 
As of today, several works have already employed ML in datasets containing WISE data. Some examples include automatic classification of sources \citep{Kurcz2016, Clarke2020, Zhao2023, Zeraatgari2024}; identification of young stellar objects \citep{Marton2019}, galaxies \citep{Krakowski2016} or active galactic nuclei \citep{Shu2019, Retana2020}; detection of debris disks candidates \citep{Nguyen2018}; unsupervised clusterization techniques to differentiate types of spiral galaxies \citep{GuoX2022}. In one study by \cite{Dobbels2020}, the authors used the ultraviolet to mid-infrared part of the SED of galaxies to predict their far-infrared emission across \textit{Herschel} bands using neural networks. We want to highlight that here (and in what follows) ``predict'' is used in the statistical sense of guessing the value of a random variable and not foreseeing the future.

In a similar approach, in this paper we exploit the extraordinary capabilities of ML techniques to find hidden relationships between different datasets to predict mid-infrared fluxes at the same wavelengths as the \textit{Spitzer} bands MIPS1 (24 $\mu$m) and IRAC4 (8 $\mu$m) from WISE variables. 
We show that this allows us to  reach a good compromise between WISE and \textit{Spitzer} capabilities, taking advantage of the large coverage of the former and the better spatial resolution and sensitivity of the latter. 
In section 2 we describe our sample and data, while section 3 details the ML procedure (algorithm and models). 
The results, including an application to the study of SEDs in a well-studied open cluster, are presented and discussed in Section 4. 
Finally, we wrap up with a summary and the concluding remarks in Section 5.

%
%--------------------------Methods1: data-----------------------

\section{\label{sec:sample}Description of the sample}

 In our attempt to make a good comparison between WISE and \textit{Spitzer} and in order to obtain consistent and trustful predictions with ML -- that is, to avoid the ``garbage in, garbage out'' problem \citep{GeigerR21} -- we need reliable input data. But, at the same time, we want a data set that is not too large so we can process it within a reasonable time. 
 With the aim of using it for further studies (Fonseca-Bonilla et al.~in prep.
 \footnote{We are developing a detailed study of warm infrared excesses that exploits the predicted fluxes obtained with our method to characterize the distribution and evolution of circumstellar disks in open clusters with high quality data.}), we have chosen the largest and most updated sample of members of a census of high quality 3530 open clusters by \citet{Hunt2024} from {\em Gaia} DR3 \citep{GaiaCollaboration2018}.
 Following the criteria in \citet{Hunt2023} and for the purpose of the ML training, we decided to discard membership probabilities of less than 50\%.
 The remaining open cluster members comprise our HR24 sample.
 To obtain the infrared data of the objects, we selected the \textit{Spitzer} Enhanced Imaging Products (SEIP) Source List CR3\footnote{\url{https://irsa.ipac.caltech.edu/data/SPITZER/Enhanced/SEIP/overview.html}} as our reference catalog, due to the high reliability of its \textit{Spitzer} data (a $10\sigma$ level of signal-to-noise ratio in at least two channels for galactic sources).
 This data set includes both \textit{Spitzer} and WISE data (fluxes and uncertainties) at several bands, along with reliability and quality flags from the allWISE Release \citep{Cutri2013} positional counterparts.
 This guarantees a consistent collection of sources with \textit{Spitzer} and WISE data with which to train and optimize the ML models, as well as a collection of sources with exclusively WISE data for which to predict mid-infrared fluxes. 
 It is necessary to take into account the consideration made in the SEIP Explanatory Supplement\footnote{\url{https://irsa.ipac.caltech.edu/data/SPITZER/Enhanced/SEIP/docs/seip_explanatory_supplement_v3.pdf}} about the completeness of its Source List, bearing in mind it was specifically designed to prioritize reliability over completeness.
 
 A query in the SEIP Source List at 2 arcsec with the coordinates of the 444219 stars of the HR24 sample resulted in 26068 positive matches with a nonzero value in the WISE W4 band at 22 $\mu$m, hereafter our ``unclean'' sample. 
 The condition of a nonzero W4 value is sufficiently restrictive to automatically guarantee nonzero values in W1, W2, and W3  (because to predict fluxes at 8 $\mu$m and 24 $\mu$m  we need values in all WISE bands and the rest of variables).
 In addition, this restriction to nonzero values in W4, addresses the interest of studying fluxes at this wavelength (e.g., for studies of warm infrared excesses) and also allows us to inspect the high discrepancies between W4 and its \textit{Spitzer} counterpart channel MIPS1 at 24 $\mu$m.
 Since WISE photometric measurements are optimized for point sources, we went a step further in terms of data reliability by setting in the query ext\_flg=0.0. 
 This constraint ensured we were looking at a point source or an object not superimposed on a 2MASS Extended Source Catalog object. 
 To avoid a spurious detection or contamination (from diffraction spikes, persistence, halo and ghost artifacts) in all WISE bands, we imposed cc\_flags='0000'.
 These two conditions left us with a ``clean'' sample of 9231 objects. 

\begin{figure}
\resizebox{\hsize}{!}{\includegraphics[width=0.9\linewidth]{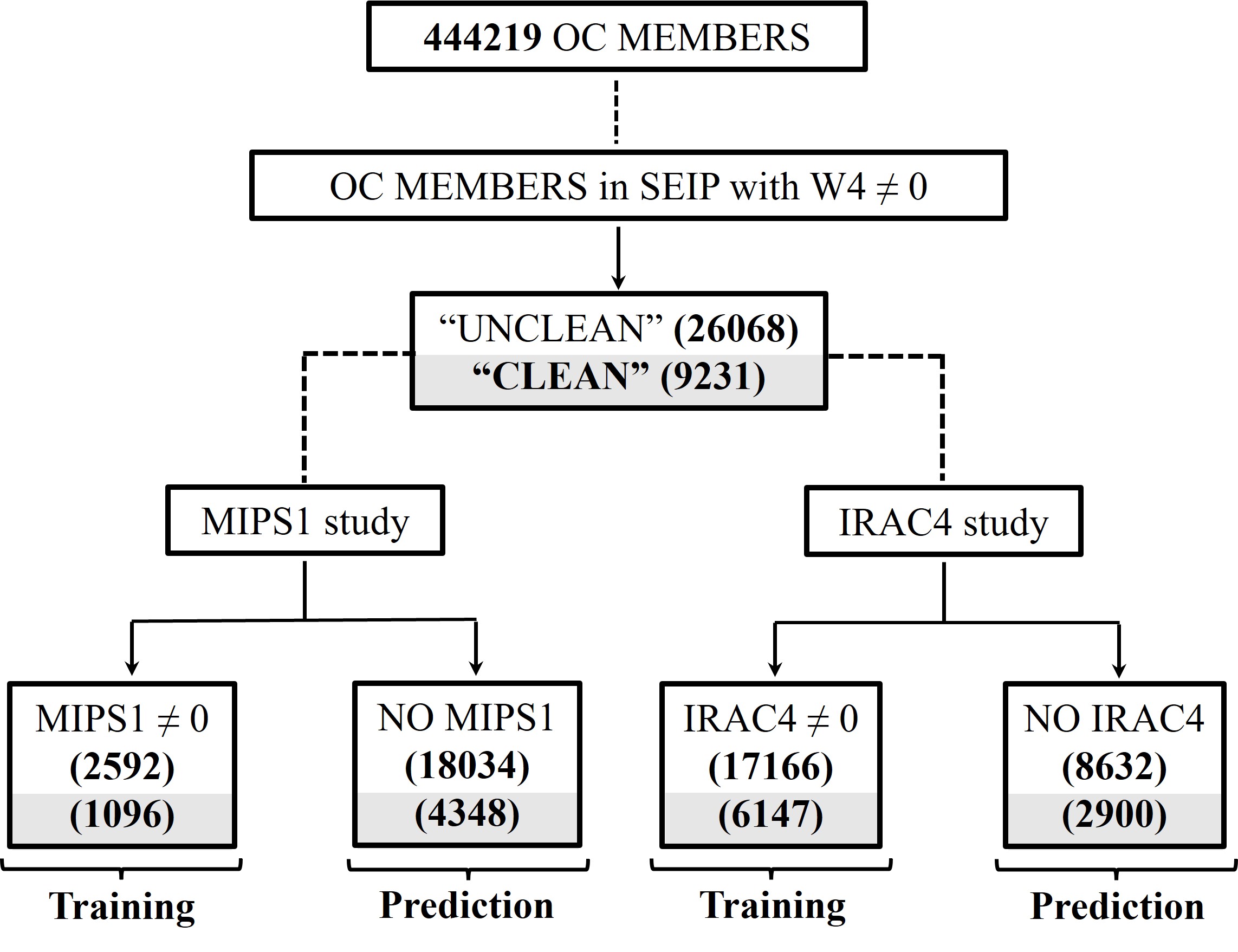}}
\caption{Data structure scheme for the ML studies in this paper. The numbers in gray shadowed boxes correspond to the clean datasets. OC stands for open cluster.}
\label{fig:data_scheme}
\end{figure}

\begin{figure}[tbhp]
\centering
\resizebox{\hsize}{!}{\includegraphics[width=0.9\linewidth]{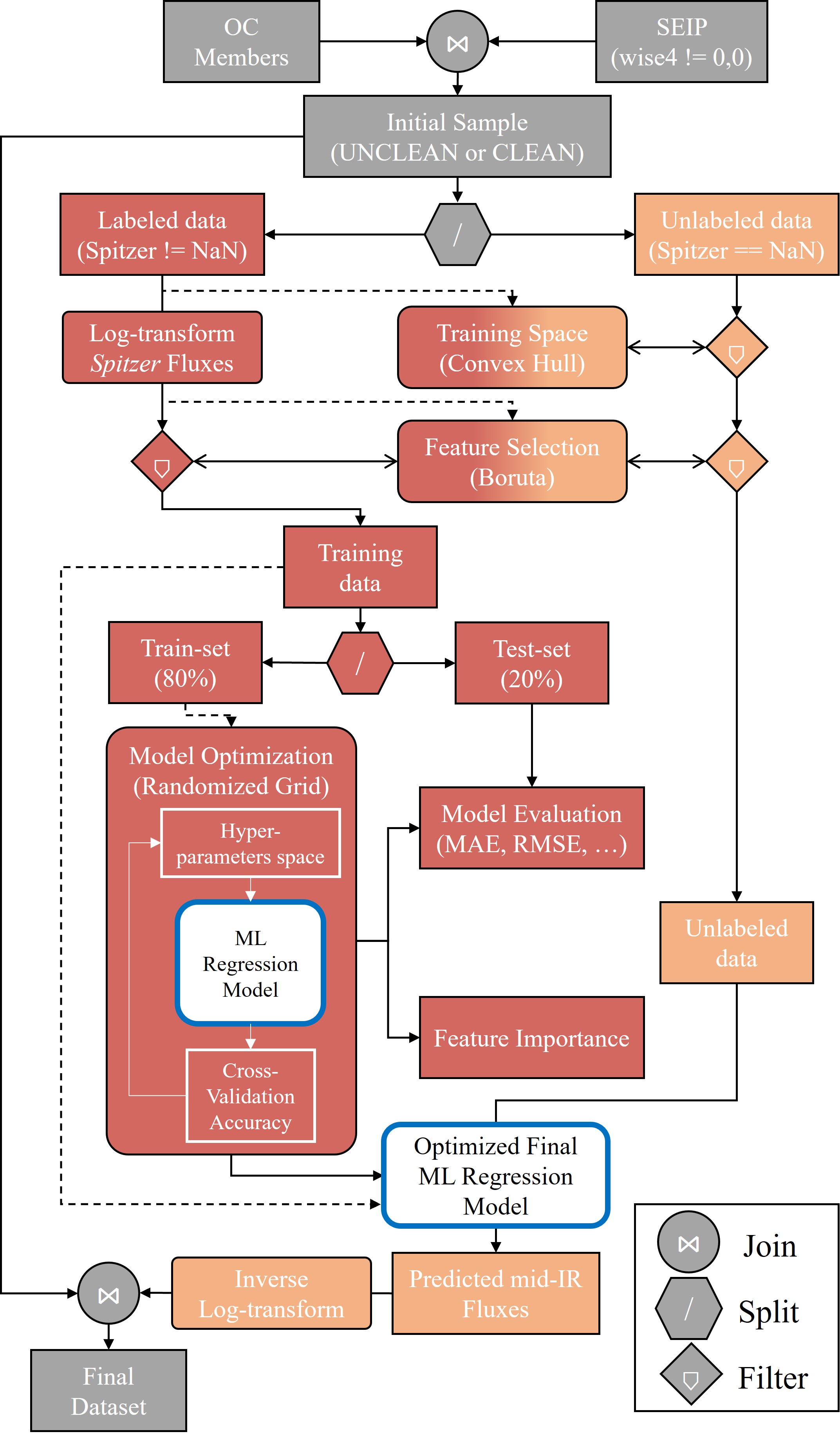}}
\caption{Workflow detailing the data management and preprocessing steps, as well as the ML model optimization, training, and evaluation procedures, to predict mid-infrared fluxes from WISE variables. Squared boxes represent datasets, while the ML models, algorithms, and other mathematical operations are indicated with rounded boxes.  The actual ML regression model (rounded boxes outline in blue) are displayed in more detail in  Fig.~\ref{fig:ML_scheme}. A dashed arrow-connector between a dataset and an algorithm signifies that the former has been used to train the latter. A double-headed arrow connecting a filter hub and an algorithm signifies ``read filtering conditions for rows or columns and apply to dataset arriving to the hub''.
}
\label{fig:data_workflow}
\end{figure}

 It is worth highlighting that due to observational issues and also the strict reliability criteria of the SEIP Source List, not all sources with flux measurements in W4 have a MIPS1 counterpart. 
 The same is true for W3 and IRAC4 (8 $\mu$m) pairs.
This disparity in data availability is displayed in Fig.~\ref{fig:data_scheme}, which shows a scheme of the database design, its structure, and the number of objects included in both unclean and clean (gray shadowed) samples.
To predict fluxes at MIPS1 band (24 $\mu$m) (left branch of the diagram), we use the subsample with nonzero values in MIPS1 for tuning and optimizing the ML models, as well as for evaluating their performance. Subsequently, the trained ML models predict fluxes at 24 $\mu$m for the rest of the sources without values in MIPS1, as explained in Section~\ref{sec:ML} and sketched in the workflow displayed in Fig.~\ref{fig:data_workflow}. 
Notice that for the MIPS1 ML studies, $\sim$ 10\% of the sources have a nonzero value in the corresponding band, evidencing the necessity of carrying out the present study.
It works similarly to predict fluxes at IRAC4 band (8 $\mu$m) from WISE variables (IRAC4 study, right branch), but in this case only a third of the sources lack a value in the \textit{Spitzer} band.

%%----------------one column figure: data structure ----------------

Another peculiarity that draws the attention in Fig.~\ref{fig:data_scheme} is the mismatch between the total number of sources in the clean and unclean datasets, and the sum of sources with \textit{Spitzer} flux densities different from zero with those without flux density values. 
For instance, we can see that for the MIPS1 study with the clean dataset, there are 1096 sources with MIPS1$\neq0$ and 4348 without MIPS1 values, whose sum is significantly smaller than the 9231 sources included in this dataset.
This loss of data comes from a restriction imposed by the ML models, which excel at interpolating (making predictions inside the training space) but, in many cases, perform poorly when extrapolating (making predictions outside the training space).
To avoid this problem, we removed from the datasets without \textit{Spitzer} fluxes all sources whose WISE parameters were outside the training space.
In particular, we looked for the convex hull (i.e., the {\em envelope}) of the training space as defined exclusively by W1 to W4, located the sources with those fluxes outside the convex hull, and discarded them.
While this step reduced the data availability, it ensured that the predictions are even more reliable.

Concerning the different variables (or features) involved in our study, we just selected those of numerical nature related with the fluxes and their qualities for all the WISE bands: WISE fluxes at the four available channels (wise\#), the reduced $\chi^2$ for all the WISE bands (wise\#rchi2), and their frame coverage quality flags (wise\#m and wise\#nm). In the case of \textit{Spitzer} we used MIPS1 and IRAC4 fluxes and errors: m1\_f\_psf and m1\_df\_psf, i4\_f\_ap1 and i4\_df\_ap1.

%%------------------one column figure: Data and ML workflow --------------------

%
%-----------------------Methods2: ML------------------------------

\section{\label{sec:ML}Machine learning tools}

We can mathematically denote the previous datasets as $\mathcal{X}=\{x_{1,i},\dots,x_{p,i}, y_{i}\}_{i=1,\dots,n}$,  where $n$ is the number of observations (stars), $p$ is the number of random variables $X_{j=1,\dots,p}$ (WISE fluxes and flags), and $x_{j,i}$ and $y_{i}$ are, respectively, the $i$-th observation of $X_j$ and the random variable $Y$ (\textit{Spitzer} fluxes). 
In the following we refer to the variables $X_j$ either as predictors or features.
Our goal is to predict the target variable $Y$ from the set of predictor variables $X_j$ by means of a {\em regression model} $Y = m(X_1,\dots,X_p)+\epsilon$, where $m$ is the regression function and $\epsilon$ is the uncertainty associated to the prediction. 
The function $m:\mathbb{R}^p\rightarrow\mathbb{R}$ encodes the relation between $Y$ and $(X_1,\dots,X_p)$ and is in general unknown. 
Thus, learning a surrogate function that provides an estimate $\hat{m}$ of the unknown function $m$ is the target of all statistical and machine learning regression models. 
Once $\hat{m}$ is learned, one can predict $Y$ given an observation of $(X_1,\dots,X_p)$.

To find a mathematical model $\hat{m}$ that can generate accurate predictions and uncertainty intervals, several distinct and fundamental steps must be followed, as depicted in the data workflow of Fig.~\ref{fig:data_workflow}.
At a high level, it can be summarized as follows: the SEIP catalog is cross-matched with the HR24 catalog. 
The resulting [initial] sample (unclean or clean, see Sec.~\ref{sec:sample}) is split into labeled (with \textit{Spitzer} fluxes) and unlabeled (without \textit{Spitzer} fluxes) datasets. The labeled dataset is used to define the training space, and the unlabeled data points outside it are excluded (or filtered) (Sec.~\ref{sec:sample}). The labeled dataset, after the \textit{Spitzer} fluxes are log-transformed, is used again to select the features that are important for the regression problem (Sec.~\ref{sec:ML_feature_select}). The irrelevant features are removed (or filtered) from both datasets. The remaining [training] data is split into train and test sets. The train set is used to tune and optimize the ML models (Sec.~\ref{sec:model_opt}), and the test set is used to evaluate the optimized model performance. The optimized final ML regression model is retrained with the whole training data (train+test). This final ML model is then used to make predictions on the unlabeled dataset. Finally, these predictions, once an inverse log-transform is applied, are appended to the initial sample as a new feature.
The actual ML regression model is described in more detail in Sec.~\ref{sec:ML_pipeline}.
These steps, as well as other relevant algorithmic aspects, will be described in what follows.

%%------------------two column figure: ML scheme--------------------

\begin{figure*}[tbhp]
\centering
\includegraphics[width=0.8\linewidth]{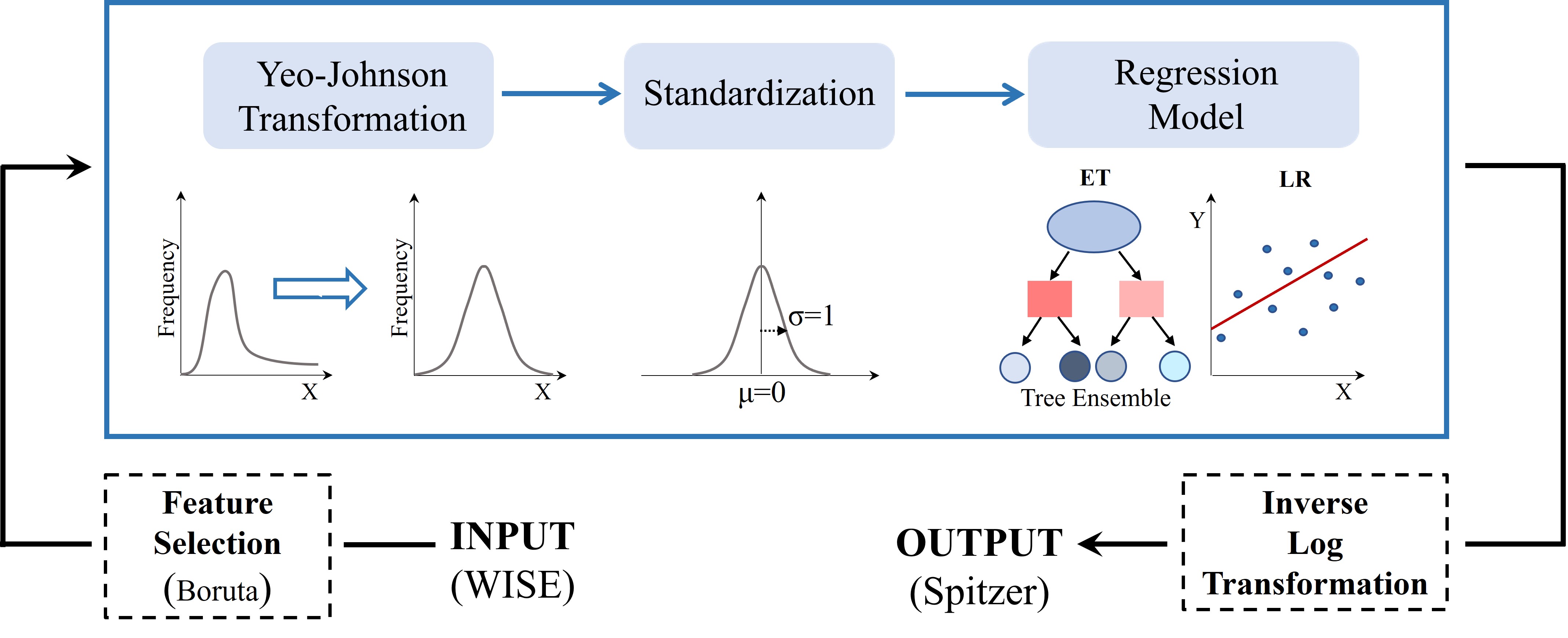}
\caption{Scheme of the steps involved in the prediction of mid-infrared fluxes from WISE features. The blue box encloses the data transformation algorithms and the regression model comprising our ML model.}
\label{fig:ML_scheme}
\end{figure*}

\subsection{\label{sec:ML_pipeline}Machine learning model pipeline}

The actual ML model consists of a ``transformation'' step and a ``regression'' step arranged in a pipeline (see Fig.~\ref{fig:ML_scheme}).

\subsubsection{Dataset transformation}

Many ML models require the predictors $X_j$ and the target $Y$ to follow normal-like distributions, but both WISE and \textit{Spitzer} data are highly right-skewed. 
Therefore, the \textit{Spitzer} fluxes were normalized applying a base 10 log transformation.
Applying this transformation to the WISE features could somewhat normalize the distribution of the fluxes, but not their quality flags.    
Accordingly, we made the WISE variables more normal-like using a Yeo-Johnson power transformation for each predictor independently (see Fig.~\ref{fig:ML_scheme}). 
This transformation relies on a data-dependent parameter $\hat{\lambda}_j$ whose optimal value is obtained maximizing the likelihood on the transformed variable \citep{Yeo2000}. 
Furthermore, some ML models are sensitive to the differences in the magnitude and spread of the predictors, assigning artificially more weight to those predictors with a larger magnitude. 
To avoid this, the Yeo-Johnson transformed predictors were subsequently  standardized by subtracting an estimated mean $\hat{\mu}_j$ and dividing by an estimated standard deviation $\hat{\sigma}_j$. 
With this preprocessing procedure, we guarantee that the predictors follow standard normal-like distributions (see Fig.~\ref{fig:ML_scheme}).
The estimates $\hat{\lambda}_j$, $\hat{\mu}_j$, and $\hat{\sigma}_j$ are chosen following a cross-validation (CV) procedure (see Section~\ref{sec:model_opt}). 
Once the predictions were obtained, an inverse log transformation was applied to yield the actual fluxes back (see Figs.~\ref{fig:data_workflow} and \ref{fig:ML_scheme}).

\subsubsection{Regression model}
\label{sec:regr_model}

Using the AutoML Python library PyCaret \citep{PyCaret}, we first made a prescreening of the performance of several regression algorithms on our datasets.
In view of the obtained results, as our main model we chose an ensemble of decision trees because they are the state-of-the-art solutions to solve regression problems \citep{Sagi2018} and because they can provide an estimation of the predictions' uncertainty, which are fundamental for astrophysical studies. 
In this kind of ML algorithms, $q$ decision trees are trained independently and their predictions are averaged to render a final prediction $\hat{y}_i$. 
A decision tree can be abstracted as follows: a) Divide the predictor space (set of possible values of $(X_1,\dots,X_p)$) into $J$ distinct non-overlapping regions $R_1,\dots,R_J$ (leaves). 
b) For each observation contained into the region $R_J$, the prediction $\hat{y}_J$ is the mean of the target values $Y_i$ for the training observations in $R_J$. 
The goal is to find regions $R_1,\dots,R_J$ that minimize the residuals between the predictions $\hat{y}_J$ and the values of $Y_i$ in each region. 
Decision trees are constructed recursively, following a binary splitting approach, from the root to the leaves, selecting for every new node the best feature and ``splitting condition'' (value of the feature) to grow the tree. 
The process begins at the bottom of the tree (root), where all observations belong to a single region. Then, in the first node, the predictor space is split, as dictated by the best feature and splitting value, into two new regions (branches). Next, the process is repeated but, instead of splitting the whole predictor space, only the observations within each of the branches are split. The splitting continues growing ever thinner branches (with fewer and fewer observations) until a stopping criterion (depth of the tree, minimum number of observations per leaf, ...) is reached, at which moment we have found all the leaves (regions $R_1,\dots,R_J$).
For regression tasks, the best feature and splitting condition for a node are the ones minimizing the average of the variances of $Y_i$ in each partition.
Using a single decision tree to make predictions is not recommended, as the prediction performance is limited and they suffer from high variance, in the sense that a small change in the data can cause a large change in the final estimated tree \citep{HastieBook09}. 
However, by aggregating many decision trees, i.e, using an ensemble of independently trained trees, the predictive performance of this kind of algorithms can be substantially improved, and the variance of the model reduced.
Among the existing tree ensemble models, we resorted to an extremely randomized trees (ET) algorithm, which is similar to the well-known random forests \citep{Breiman2001}, but with an extra component of randomization \citep{Geurts2006}. 
As in random forests, in ET each tree is built using only a random subset of predictors $X_j$, but the features and splitting condition for each node are also set randomly.
ET has several hyperparameters that must be tuned to improve the prediction capability.
The main ones are the number of trees in the ensemble ($q$), the depth of the trees (maximum number of splits), the minimum number of observations in a partition in order to continue subdividing the tree, or the minimum number of observations in the leaves.
We refer interested readers to the documentation of \texttt{scikit-learn} implementation of ET (\texttt{ExtraTreesRegressor}) for a full description of all the hyperparameters\footnote{\url{https://scikit-learn.org/stable/modules/generated/sklearn.ensemble.ExtraTreesRegressor.html}}.
As baseline model, we chose a basic multidimensional linear regression (LR) model that assumes $Y = \beta_0+\beta_1X_1 + \dots +\beta_pX_p+\epsilon$, where $\beta_0$ is the intercept and $\beta_1,\dots,\beta_p$ are the slopes.

\subsubsection{Uncertainties of predicted mid-infrared fluxes}
\label{sec:predicted_err}

To estimate the uncertainty intervals of the individual predictions, we leveraged the ensemble nature of ET. 
As explained above, each of the $q$ trees of the ET gives a prediction for the $i$-th observation, and the final prediction $\hat{y}_i$ is the mean of those $q$ predictions. 
Grounded on the Central Limit Theorem, the prediction uncertainty ($3\sigma$ confidence level) was obtained from the standard error of the mean as $\sigma_{\hat{y}_i}=3\sigma_i/\sqrt{q}$, where $\sigma_i$ is the standard deviation of the $q$ predictions for the $i$-th observation. 
The standard error was obtained for the log transformed fluxes. 
Accordingly, the lower and upper bounds of the predicted ML fluxes were obtained as $10^{\hat{y}_i-\sigma_{\hat{y}_i}}$ and $10^{\hat{y}_i+\sigma_{\hat{y}_i}}$, respectively. 

There are more sophisticated methods to infer the uncertainty: using a resampling strategy such as bootstrap \citep{EfronBook93}; resorting to more advanced probabilistic regression algorithms such as natural gradient boosting \citep{duan2020}, bayesian neural networks \citep{NealBook2012, Hunt2023}, or mixture density networks \citep{BishopCM94}; or estimating the uncertainty as a parameter using a specifically defined loss function \citep{PearceT2018, Dobbels2020}. Unfortunately, all these algorithms are ``data-hungry,'' computationally expensive, and/or increase the problem complexity. Given that our current dataset is small (less than 1000 training instances), we decided to use the standard error for the uncertainty estimation, as it provided the best trade-off between accuracy and simplicity.

\subsection{\label{sec:ML_evaluation}Machine learning model optimization and evaluation metrics}
    
\subsubsection{Model selection and optimization}
\label{sec:model_opt}

As we have just seen in Section~\ref{sec:ML_pipeline} the ML pipeline is built from the concatenation of data transformation algorithms and a regression model.
In the remaining of the paper, with LR, ET, or ML models we refer to the whole pipeline, not just the regression step.
    
The parameters of the transformation step ($\hat{\lambda}_j$, $\hat{\mu}_j$, and $\hat{\sigma}_j$) and the hyperparameters of the regression step (number of trees, tree depth, etc...) must be optimized to render the best predictions while avoiding overfitting (i.e., learning the training data by heart instead of inferring the regression function $\hat{m}$). 
With that aim, the whole dataset was randomly split into train and test datasets, with an 80\% of observations assigned to the former. 
The train dataset was used to select the best hyperparameters, while the test dataset was used to independently assess the performance of the optimal model against new observations (Fig.~\ref{fig:data_workflow}). 
The best hyperparameters were found using $k$-fold CV and a randomized grid search. 
In the CV step \citep{HastieBook09}, the train dataset is randomly split into $k$ subsets (folds). 
The transformation parameters ($\hat{\lambda}_j$, $\hat{\mu}_j$, and $\hat{\sigma}_j$) and the regression model --with a particular configuration of hyperparameters--, are fitted using all observations from $k-1$ folds. The prediction performance (see Goodness of fit metrics below) is then evaluated on the observations of the left-out fold (after they have been transformed according to the fitted parameters $\hat{\lambda}_j$, $\hat{\mu}_j$, and $\hat{\sigma}_j$). 
This procedure is repeated with all possible combinations of $k-1$ folds, rendering a total of $k$ independent evaluations of the ML model performance. 
The overall predictive capability of the ML model is given by the mean of the $k$ goodness of fit metrics values. 
The best hyperparameters are those resulting in the best mean metrics.    
The hyperparameters configurations evaluated with the CV procedure were chosen following a random grid search \citep{BergstraJ2012}. 
In this search strategy, for each configuration, the values of the hyperparameters are randomly chosen among all possible combinations.
For example, for ET, the number of trees in the ensemble can be any integer between 10 and 500.
For the same number of trials, the randomized search finds better combinations of hyperparameters than the regular grid search, which divides the search space evenly.
For our ML study, we used 10-fold CV and 1000 configurations in the hyperparameters search.

Finally, as we sketch in Fig.~\ref{fig:data_workflow}, the ML model with the optimal hyperparameters was retrained with the whole dataset (train + test) to obtain the final model. This model was then used to make predictions on the samples whose target variable (\textit{Spitzer} fluxes) was unknown. These predictions can then be used for further astrophysical studies.

\subsubsection{Goodness of fit metrics}

To quantitatively evaluate the prediction performance of the ML models (goodness of fit), we used the mean absolute error (MAE), the root mean squared error (RMSE), the coefficient of determination $R^2$, and the mean absolute percentage error (MAPE). Let $y_i$ and $\hat{y}_i$ be, respectively, the true and predicted values of the target variable for the $i$-th observation, and let us define $\bar{y}=\sum_i^Ny_i/N$, with $N$ the number of observations that are evaluated. Then, the evaluation metrics are computed as follows:

\begin{equation*}
\begin{aligned}
&\textrm{MAE}=\frac{1}{N}\sum_i^N\left|y_i-\hat{y}_i\right|
&\textrm{RMSE}&=\sqrt{\frac{1}{N}\sum_i^N\left(y_i-\hat{y}_i\right)^2}\\
&R^2=1-\frac{\sum_i^N\left(y_i-\hat{y}_i\right)^2}{\sum_i^N\left(y_i-\bar{y}\right)^2}
&\textrm{MAPE}&=\frac{1}{N}\sum_i^N\frac{\left|y_i-\hat{y}_i\right|}{\left|y_i\right|}.
\end{aligned}
\end{equation*}

MAE and RMSE have dimensions and their values depend on the magnitude of the target variable. In contrast, $R^2$ and MAPE are adimensional and independent of the target magnitude.

\subsection{\label{sec:ML_features}Feature selection and importance}

\subsubsection{Feature importance}
\label{sec:ML_feature_impor}

One of the key aspects of any regression model is to be insightful in relation to which is the ``importance'' of each feature. This is, its predictive power or how relevant it is to make predictions.
In the LR model, for instance, the feature importance can be obtained from the slopes $\beta_j$. 
The larger the slope is, the larger the importance of its feature is.
For the ET model, though, a different approach is followed. 
In the \texttt{scikit-learn} ET implementation the importance is proportional to the number of times a feature is selected for a node, with the nodes closer to the root having more weight than those closer to the leaves, and averaged over all trees of the ensemble \citep{BreimanBook84}.
In order to make meaningful comparisons between models, the feature importances are usually normalized so that they add up to one. 

\subsubsection{Feature selection}
\label{sec:ML_feature_select}

Discarding those features that are irrelevant, unimportant or even redundant for the regression problem, helps to reduce the dimensionality of the problem and improves the outcome and accuracy of the ML models \citep{Chandrashekar2014}.
With that aim, before the ML model optimization and training steps, we employed Boruta \citep{KursaMB2010}, an iterative feature selection algorithm that removes one by one the unimportant features (Figs.~\ref{fig:data_workflow} and \ref{fig:ML_scheme}). 
It consists on repeating the following steps a number of times or until there are no more irrelevant features.
Starting from the standardized dataset $\tilde{\mathcal{X}}=\{\tilde{x}_{1,i},\dots,\tilde{x}_{p,i}, \tilde{y}_{i}\}_{i=1,\dots,n}$, another one is created by randomly shuffling each feature $X_j$. 
These ``shadow'' features are joined with the original dataset, obtaining a new one, $\mathcal{X}_S=\{\tilde{x}_{1,i},\dots,\tilde{x}_{p,i},x_{1,i}^s,\dots,x_{p,i}^s, \tilde{y}_{i}\}_{i=1,\dots,n}$, with twice as many features.
An ML regression model is then fitted to $\mathcal{X}_S$, and the importance of each feature is computed.
If the importance of an original feature is larger than those of the shadow ones, then it is considered relevant for the regression task and is assigned a {``hit''}. 
Given the binary outcome of each independent experiment (hit or no-hit), the probability of getting $k$ hits after $m$ trials follows a binomial distribution.
Therefore, those features with probabilities below the confidence level $\alpha$ after $m$ trials are deemed irrelevant and, consequently, are removed from the dataset.
Then the shuffling, hit assignment, and statistical test process starts again but without the discarded features. 
For this study, we used as ML model the boosted tree algorithm \texttt{LightGBM} \citep{KeG2017}, and set $m=100$ and $\alpha = 0.05$.

%
%--------------------------Results-------------------------------------

%%-----------------------two column figure----------------------------

\begin{figure*}[tbhp]
\centering
\includegraphics[width=\linewidth]{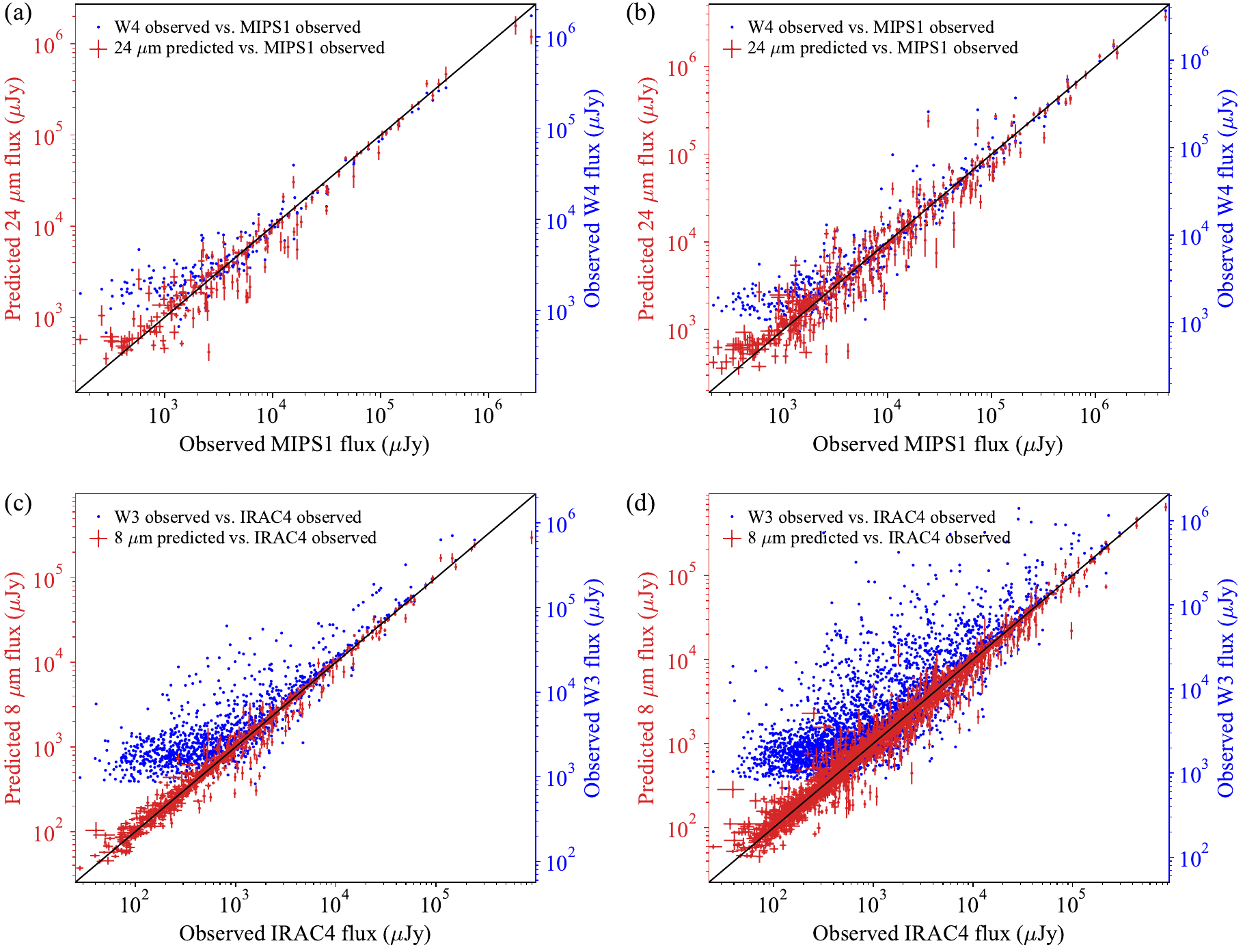}
\caption{Predicted mid-infrared fluxes (red points, left axes) and observed WISE fluxes (blue points, right axes) vs.~observed \textit{Spitzer} fluxes. All panels display the fluxes for the stars in the corresponding test set. (a) and (b) display the values of MIPS1 (24 $\mu$m) and W4 fluxes, while (c) and (d) the values of IRAC4 (8 $\mu$m)  and W3 fluxes. (a) and (c) show the fluxes of the clean test sets, while (b) and (d) do the same for the unclean test sets. The error bars for the observed WISE data (blue points) have not been included to avoid clogging the plot. 
Diagonal black lines have been included in all plots to serve as reference and facilitate the comparison.
Notice that the limits of both Y axes in each plot, predicted (left) and observed (right), are dissimilar, as they represent fluxes at different wavelengths.
In fact, they differ in scale factors of $(24/22)^2$ (for a and b) and $(12/8)^2$ (for c and d), precisely the ratio of fluxes at different wavelengths expected for a black body model.
The limits in both axes are chosen so that both clouds of points (blue and red) fall in the diagonal in case both satellites provided perfectly accurate measurements. 
}
\label{fig:ML_predictions}
\end{figure*}

\section{\label{sec:results}Results and discussion}

As mentioned in the Introduction, WISE data, in comparison with \textit{Spitzer} data, can be less sensitive and more confused, either by overlapping close-by sources or compact structures in the sky background. Therefore, a given sample of objects, for example, debris disks, can be contaminated by other type of unresolved objects such as, for instance, infrared galaxies. 
This fact stands out when comparing \textit{Spitzer} and WISE data, as can be seen in Fig.~\ref{fig:ML_predictions}. 
The correlation between MIPS1 and W4 observed fluxes for a selection of sources in our clean and unclean samples is shown in Figs.~\ref{fig:ML_predictions}a and b, respectively.
While for MIPS1 fluxes above $\sim10^3$ $\mu$Jy there is a good correlation between the values in both bands, below that value the fluxes differ in up to an order of magnitude. A similar behavior can be observed when comparing IRAC4 and W3 observed fluxes in the same samples (see Figs.~\ref{fig:ML_predictions}c and d). 
One possible solution to resolve this discrepancy is to make a thorough selection of the objects in the sample attending to quality criteria, but this ends up in a huge loss of data (see Appendix \ref{appendix_1}). 
Here we propose an alternative approach, taking advantage of the powerful prediction capabilities of ML techniques and algorithms.

\subsection{Machine learning models predictions and goodness of fit metrics}

One of the several applications of ML is the treatment of correlated datasets in which their specific relationships are not well known  \citep{HastieBook09, LindholmBook22}. 
\textit{Spitzer} and WISE observed the sky at similar wavelengths and, in general, the tendencies between the different bands of each satellite are expected to be similar for each target, making it an ideal case scenario to use ML. 
In other words, one could possibly find a regression function (i.e., ML model) relating the WISE fluxes (and their quality flags) to those of \textit{Spitzer}.
As we previously said in Section~\ref{sec:regr_model}, we chose ET as our main ML model, since it was the one that gave the best results across samples among the many other algorithms that we preliminarily tested. 
Besides, it provides an estimation of the prediction uncertainties (see Section~\ref{sec:predicted_err}), which is crucial for any astrophysical study.
For comparison and to test the performance of our method, we also made a trial with a basic LR model.

Fig.~\ref{fig:ML_predictions} displays the predicted mid-infrared fluxes at MIPS1 and IRAC4 wavelengths obtained with the ET models trained with either the clean or the unclean datasets. For a better comparison, the predicted fluxes (red points) are superimposed with the observed WISE ones (blue points).
We want to emphasize that all points in Fig.~\ref{fig:ML_predictions} --predicted mid-infrared fluxes and observed WISE ones-- belong just to the sources in the test sets, which account for the 20\% of the corresponding datasets and have not been used in the training (or optimization) phase of the ML models.
Accordingly, they serve as a test bed of their performance for observations that do not have \textit{Spitzer} fluxes.
An ideal prediction would result in all red points to lay perfectly in the bisectors of Fig.~\ref{fig:ML_predictions} (diagonal black lines), as it would mean that the predictions are exactly equal to the observed values.
Attending to this criterion, we can conclude that the predictions of our ET models are rather good, for both mid-infrared bands, and for the clean and unclean datasets.
Remarkably, the predicted mid-infrared fluxes fall close to the diagonals at all ranges of the observed \textit{Spitzer} fluxes, meaning that our ET models are very accurate even for the weakest fluxes.
This single fact is the most important result of our study and highlights the relevance of our ML approach, as it opens the door to utilizing the full capabilities of WISE (specially its wide coverage) even for targets whose W3 or W4 fluxes fall below $\sim10^3$ $\mu$Jy, where this satellite shows the strongest overestimations.
Furthermore, the predicted mid-infrared fluxes lack the clear and strong dispersion that can be noticed in the observed WISE fluxes, mostly in the plots corresponding to our unclean data (Figs.~\ref{fig:ML_predictions}b and d). 

\begin{figure*}[tbhp]
\centering
\includegraphics[width=\linewidth]{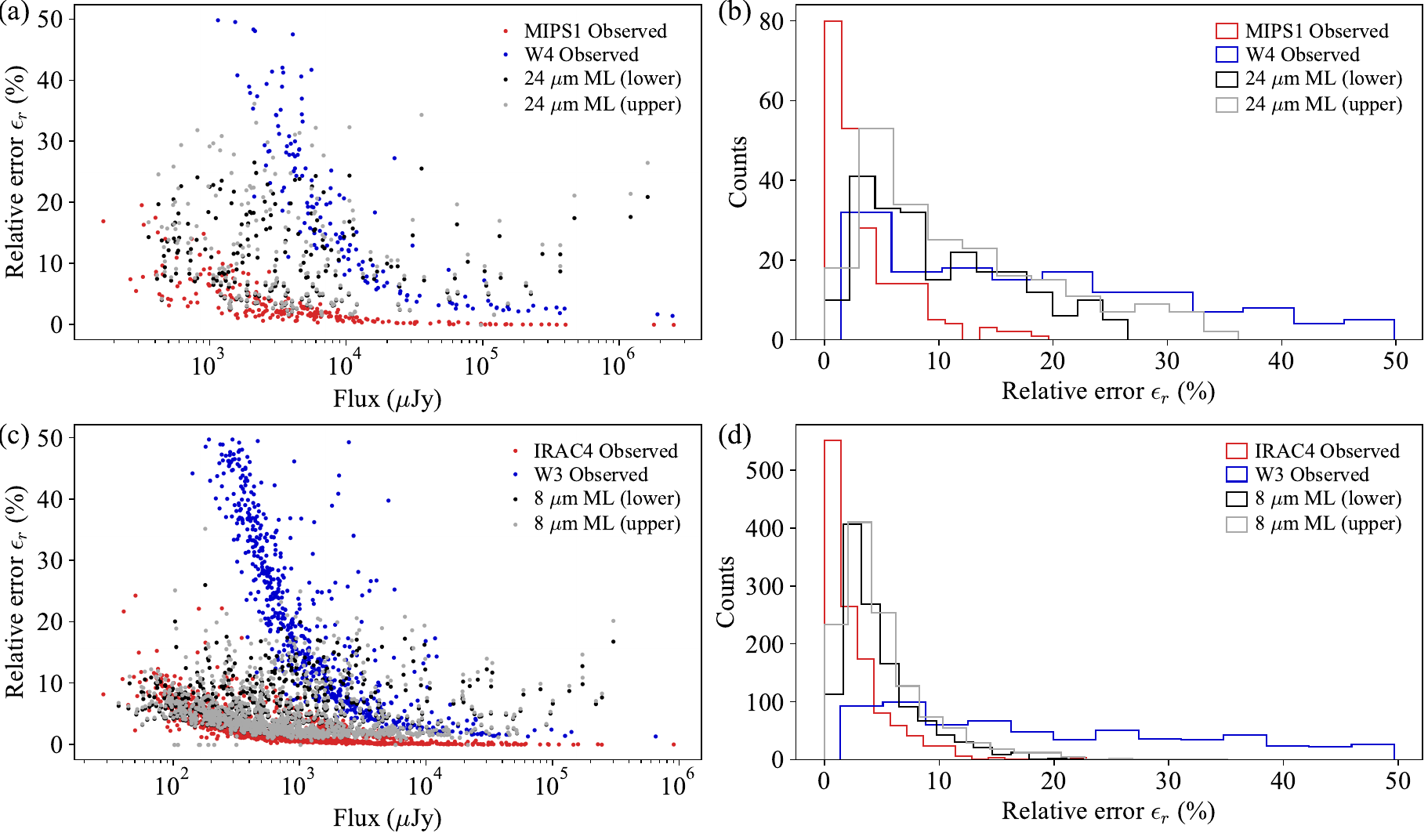}
\caption{Calibration of ML prediction errors for samples in Figs.~\ref{fig:ML_predictions}a and c. a) Relative error $\epsilon_r$ as a function of the corresponding flux for MIPS1 observations (red points), W4 observations (blue points), and the lower (black points) and upper (gray points) prediction intervals for the ML predictions at 24 $\mu m$. b) Distribution of the relative errors in a) (same color code). c) Relative error $\epsilon_r$ as a function of the corresponding flux for IRAC4 observations (red points), W3 observations (blue points), and the lower (black points) and upper (gray points) prediction intervals for the ML predictions at 8 $\mu m$. d) Distribution of the relative errors in c) (same color code).
}
\label{fig:error_calibration}
\end{figure*}

In addition, the single prediction uncertainties (vertical error bars in Fig.~\ref{fig:ML_predictions}), of a $3\sigma$ confidence level, are highly constrained and display a nearly constant relative error (uncertainty over value ratio, $\epsilon_r=100\Delta y/y$), showing a slight increase toward the weakest predicted fluxes, much like the observed \textit{Spitzer} and WISE fluxes (Figs.~\ref{fig:error_calibration}a and c). 
While at large predicted fluxes the uncertainties are significantly larger than those of the \textit{Spitzer} observed fluxes, at the weakest fluxes the former are of the same order of magnitude, even in some instances smaller, than the latter.
At the same time, the relative errors of the predicted fluxes are smaller than those of the observed WISE fluxes, specially at the weakest magnitudes. 
Notice that at this range WISE contains numerous upper limits.
But, importantly, most relative errors for the predicted fluxes are below 10\%, as clearly seen in the histograms of Figs.~\ref{fig:error_calibration}b and d.

The results in {Figs.~\ref{fig:ML_predictions} and \ref{fig:error_calibration}} suggest that the mid-infrared fluxes, predicted from WISE data using our ML approach, are quite reliable, showing both high accuracy (low bias) and high precision (low variance).
We can support this claim by looking at the goodness of fit metrics for all the ML models and datasets that we have analyzed in this study, which we have summarized in Table~\ref{table:fitmetrics}.
We remind the reader that values closer to zero of MAE, RMSE and MAPE indicate a better fit of the model.
The same holds true for $R^2$, but when it is close to one. 
Keeping this in mind, an exhaustive inspection of the numbers in the table tells us that the ET models optimized and trained with the different datasets provide very accurate predictions, confirming the results of Fig.~\ref{fig:ML_predictions}.
In fact, the coefficient of determination $R^2$ is, in the worst case (MIPS1 clean), already at $\sim0.94$, while the MAPE, which conveys information on the mean relative error, is in all cases below $\sim3.5\%$.
In addition, the MAE and RMSE for all ET models are very close to 0.

Nevertheless, the prediction capabilities of the ET models differ from sample to sample.
For example, 8 $\mu$m fluxes are predicted with a higher accuracy than those at 24 $\mu$m, both for the unclean and clean datasets, as testified by the goodness of fit metrics of $R^2\sim0.98$ and MAPE$\sim2\%$.
It is well known that ML models improve their prediction performance when the number of training instances increases \citep{HastieBook09}.
Hence, given that there are nearly eight times more sources with IRAC4 fluxes than with MIPS1 ones (cf.~Figs.~\ref{fig:data_scheme} and~\ref{fig:ML_predictions}), it is highly probable that the observed improvement in the prediction accuracy comes from an increased number of training data.
However, this principle is not general. For instance, the unclean datasets contain nearly three times more data than the clean ones (cf.~Figs.~\ref{fig:data_scheme} and~\ref{fig:ML_predictions}), but the goodness of fit metrics of the former are slightly worse than those of the latter.
This reduction in the prediction performance originates in the lower quality of the unclean datasets.
The inclusion of both non-point-like sources ($\textrm{ext\_flg}\neq0.0$) and spurious detections ($\textrm{cc\_flags}\neq\textrm{'0000'}$) in these datasets contributes with irrelevant information (noise) that, eventually, ruins the positive effect of an increased number of training instances.
This is a clear example of the ``garbage in, garbage out'' problem \citep{GeigerR21} that we cited in Section~\ref{sec:sample}.

%%%--------------------Table -------------------------

\begin{table}
    \caption{Goodness of fit metrics for the different ML models obtained during the ten-fold cross-validation step.}            
    \label{table:fitmetrics}      
    \centering                          
    \begin{tabular}{l ccccc}
    \hline            
    \hline
    \noalign{\smallskip}
    \multirow{2}{*}{Metric} & ML & MIPS1 & MIPS1 & IRAC4 & IRAC4  \\
            & model & clean & unclean & clean & unclean       \\
    \noalign{\smallskip}
    \hline
    \hline
    \noalign{\smallskip}
    \multirow{2}{*}{MAE$^{(a)}$} & ET & 0.106 & 0.117 & 0.048 & 0.063 \\
        & {\em LR} & {\em 0.171} & {\em 0.188} & {\em 0.091} & {\em 0.106} \\
    \multirow{2}{*}{RMSE$^{(a)}$} & ET & 0.165 & 0.179 & 0.083 & 0.110 \\
        & {\em LR} & {\em 0.238} & {\em 0.254} & {\em 0.139} & {\em 0.160} \\
    \multirow{2}{*}{$R^2$} & ET & 0.940 & 0.945 & 0.983 & 0.973 \\
        & {\em LR} & {\em 0.873} & {\em 0.890} & {\em 0.952} & {\em 0.944} \\
    \multirow{2}{*}{MAPE$^{(a)}$} & ET & 0.031 & 0.033 & 0.018 & 0.022 \\
        & {\em LR} & {\em 0.048} & {\em 0.051} & {\em 0.033} & {\em 0.036} \\
    \noalign{\smallskip}
    \hline                                   
    \end{tabular}
    \tablefoot{
    \tablefoottext{a}{The values corresponding to these metrics refer to log transformed fluxes, not actual fluxes.}}
\end{table}
%%%

To put the goodness of fit metrics of the ET models into perspective, it is illustrative to compare them with those obtained with our baseline model, a basic LR (see Section~\ref{sec:regr_model}).
While the goodness of fit metrics for the LR models are surprisingly good, it is clear that the ET models are far better (Table~\ref{table:fitmetrics}).
This fact comes as no surprise, as the ET models are intrinsically nonlinear models that are, in principle, capable of capturing the complex nonlinear relationships expected to exist between the WISE variables and the \textit{Spitzer} fluxes.
In any case, the LR models show a similar behavior to the ET models in what the metrics of the different datasets refers.
The LR model makes better predictions for 8 $\mu$m fluxes than for the 24 $\mu$m ones, and they are better when using the clean datasets for training.

\begin{figure}[tbhp]
\centering
\includegraphics[width=0.9\linewidth]{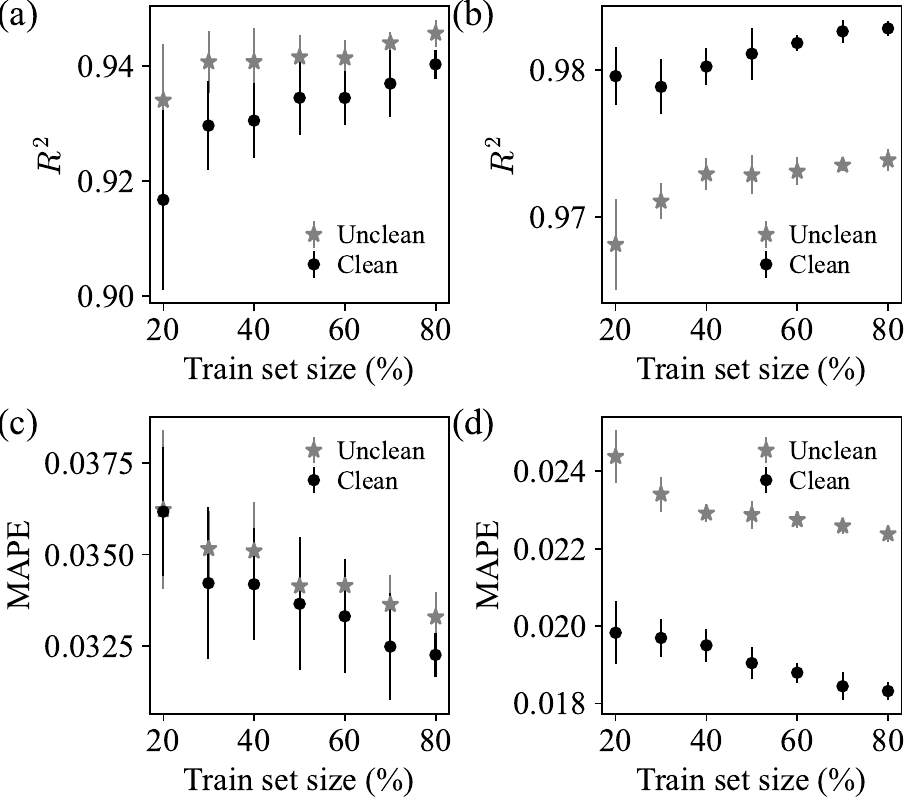}
\caption{Effect of train set size on the ET models' predictive performance: Coefficient of determination $R^2$ and MAPE as a function of the percentage of training instances used to train the models for both the clean and unclean samples. a) and c) correspond to the predictions of fluxes at $24\; \mu$m, while b) and d) to those at $8\; \mu$m. The symbols and error bars represent the mean and standard deviation of the ten-fold cross-validated metrics obtained for ten independent random samplings.}
\label{fig:learning_rate}
\end{figure}

Notice that the goodness of fit metrics collected in Table~\ref{table:fitmetrics} correspond to the average metrics obtained during the 10-fold cross-validation step (see Section~\ref{sec:model_opt}), and thus serve as an objective evaluation of the models' performance against new observations (targets without \textit{Spitzer} fluxes).
In addition, those values can be understood as a ``pessimistic'' estimation of the prediction performance of the proposed ML models. The reason is that those metrics were obtained for our ET and LR models trained only with the train dataset, that accounts for the 80\% of the points in the whole sample. Once the final ML models are retrained with the whole dataset, both the goodness of fit metrics and the mid-infrared fluxes predictions are expected to be even better, as there are more training instances from which to learn hidden relationships \citep{HastieBook09}. Indeed, to test this claim, we represent in Figure~\ref{fig:learning_rate} the ``learning curves'' of our ET models. These curves clearly demonstrate how both $R^2$ and MAPE improve monotonically with the number of training instances, both for the clean and the unclean datasets and for the predictions at 8 and 24 $\mu$m.

Even with the very good fit metrics, we can still observe, specially in Figs.~\ref{fig:ML_predictions}a and b, a slight systematic error in the predictions (overestimation of the fluxes below $3\times10^3\;\mu$Jy). This is an expected behavior for nonlinear regression models that, either are not capable of capturing the exact nonlinear relationship between the features and the target variable, or because the features does not contain enough information to explain the variability of the target variable in full. Since ET models have an arbitrary level of nonlinearity, the small bias must come from a lack of information. Indeed, the ET models for the prediction of fluxes at 24 $\mu$m have coefficient of determination of $R^2\sim0.94$ (Table 2), meaning that the features selected by Boruta can explain a 94\% of the variability. The remaining variability, not capture by the ET model, must come from missing information (other features not considered in this study) or random noise. As this systematic error cannot be fully solved with our current approach, we recommend using these predictions as part of global statistical studies (so that potential biases cancel out), or be exquisitely cautious when using the individual predictions.

\begin{figure*}[tbhp]
\centering
\includegraphics[width=\linewidth]{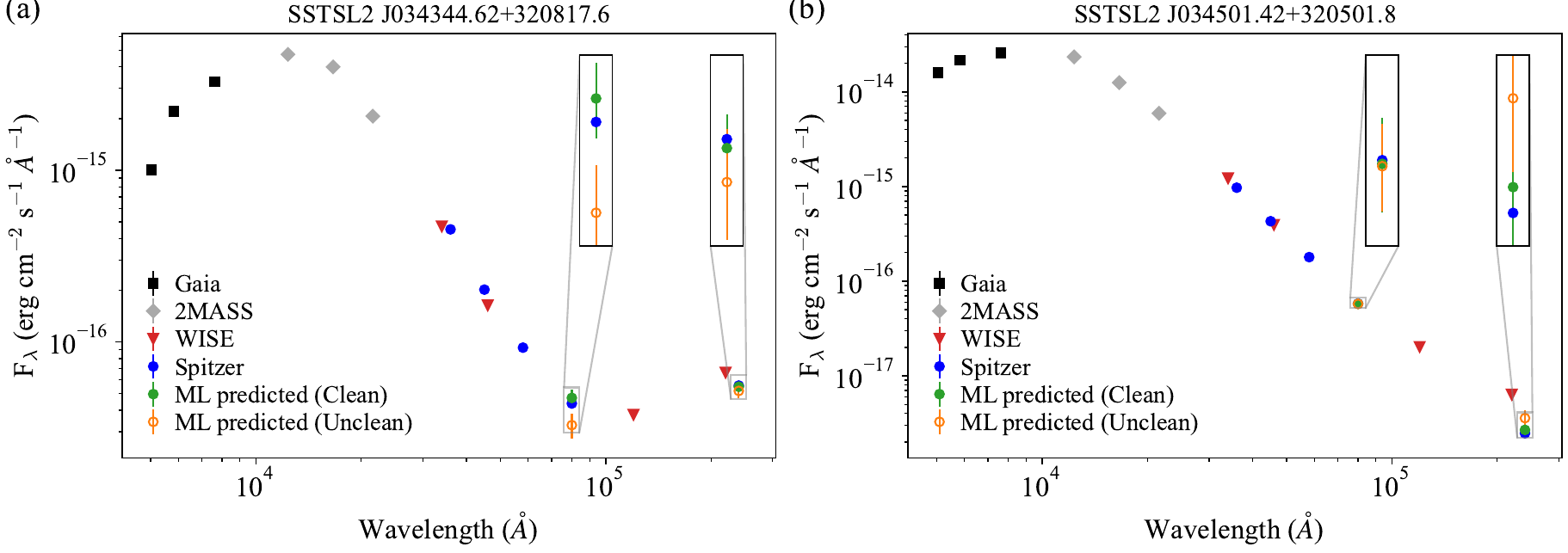}
\caption{SEDs for the sources SSTSL2 J034344.62+320817.6 (a) and SSTSL2 J034501.42+320501.8 (b) in the well-studied open cluster IC 348. The SEDs have been constructed with observed fluxes from {\em Gaia} (black squares), 2MASS (gray diamonds), WISE (red triangles), and \textit{Spitzer} (blue circles), as well as with ML predicted mid-infrared fluxes using the clean (green circles) and unclean (hollow orange circles) datasets. The insets display zoom-ins for a better comparison between the observed and predicted mid-infrared fluxes.}
\label{fig:seds_ic348}
\end{figure*}

\subsection{\label{sec:SEDs}SEDs with predicted mid-infrared fluxes: A case study in IC 348}

In order to make a visual performance test of our model, we built SEDs mixing real and predicted data for a better comparison. 
From all the open clusters in \citet{Hunt2024}, we selected the young well-studied open cluster IC 348 (see Fig.~\ref{fig:seds_ic348}) as our case study. 
Just fourteen of its members have measured values of MIPS1 and IRAC4 in SEIP with no contamination in WISE, as in our clean sample. From those fourteen, after an exploratory examination of their SEDs with VOSA \citep{Bayo2008}, we chose the two most representative cases showing discrepancies between WISE, measured \textit{Spitzer} and predicted mid-infrared fluxes (see Fig.~\ref{fig:seds_ic348}).

Although this open cluster is not included in our sample as it fails to fulfill the high quality criteria in \citet{Hunt2024}, it gives us the opportunity to perform a meaningful comparison between observed and ML predicted fluxes by using the already trained ML models. By the own nature of ensemble tree algorithms, once the model is trained, its predictions for samples that have been used in the training are almost perfectly accurate. In this case, the members in IC 348 have not been ``seen'' by the ET models, and then the fluxes predicted for them represent a sort of ``real-world scenario'' of the model performance.

We constructed each SED with SEIP \textit{Spitzer} MIPS1 and IRAC1-4, WISE and 2MASS fluxes. The infrared data was complemented with optical {\em Gaia} DR3 photometry. 
We also included our ML predicted mid-infrared fluxes. 
In both SEDs, W4 fluxes may seem almost outliers with respect to the rest of the measured data in the plot, but this would not be necessarily noticed in case of not having measured \textit{Spitzer} fluxes. 
Besides, as we have said before, as \textit{Spitzer} was designed to perform pointed observations and already reached the end of its mission, the usual scenario would be to not have its data for comparison. Furthermore, WISE measurements do not have to be of low quality to show this kind of discrepancy. 
In fact, both SSTSL2 J034344.62+320817.6 (Fig.~\ref{fig:seds_ic348}a) and SSTSL2 J034501.42+320501.8 (Fig.~\ref{fig:seds_ic348}b) have 3<w4snr<10. 
Also, in both cases, there is no indication of variability in SEIP data. In contrast, our predicted mid-infrared fluxes are in good agreement with the measured ones and fall within the expected value range, as can be seen in the insets of both SEDs. 
The other noticeable and expected result that stands out when inspecting the SEDs, is that our clean predicted fluxes provide a better fit than the unclean ones, in line with the metrics summarized in Table~\ref{table:fitmetrics}.

The importance of having good values of the flux is mostly revealed when dealing with quantitative studies, as happens in the case of looking for excesses in the infrared emission of stars that might be indicative of the presence of circumstellar disks (a detailed study with ML predicted mid-infrared fluxes applied to the characterization of disks can be found in Fonseca-Bonilla et al. in prep.). 
An overestimation in the measured flux, something that seems frequent at low WISE fluxes (Figs.~\ref{fig:ML_predictions} and \ref{fig:app}), may result in a ``false positive'' excess and lead to the wrong conclusion that a disk candidate has been found.
We believe some publications studying infrared excesses using WISE data might have been affected by this bias, as there are some disagreements with similar \textit{Spitzer} studies (see discussion in \citet{Patel2014}). Any other analysis involving not only quantitative data but the shape of SEDs (built with WISE fluxes) could also compromise the results obtained. Hence the importance of having proofed reliable data, such as our ML predicted mid-infrared fluxes. Other scientific cases of WISE studies that could benefit from better data at similar mid-infrared wavelengths and strict criteria can be found in \citet{Kurcz2016} (low level of detections at wavelengths longer than W2 challenges the reliability of automatic classification of sources), \citet{Dennihy2020} (confusion and care must be taken when selecting large sample of WISE-selected infrared excess) or \citet{Sedgwick2022} and \citet{Suazo2022} (where improving WISE detection limits would lead to better and even more precise results when searching for giant planets in the outer Solar System or Dyson spheres in the Milky Way).

\subsection{Discussion on the relevance of WISE features to predict mid-infrared fluxes}

The main goal of this study is to find an alternative method for working with WISE data by taking full advantage of \textit{Spitzer} capabilities and without loosing data in the process. 
As our analysis is specifically based in ML techniques, in what follows we will focus on the details of the feature selection and importance with the aim of getting to the bottom of the ML model. 
This will help us understand which are the most relevant WISE features to predict mid-infrared fluxes.

\begin{figure*}[tbhp]
\centering
\includegraphics[width=\linewidth]{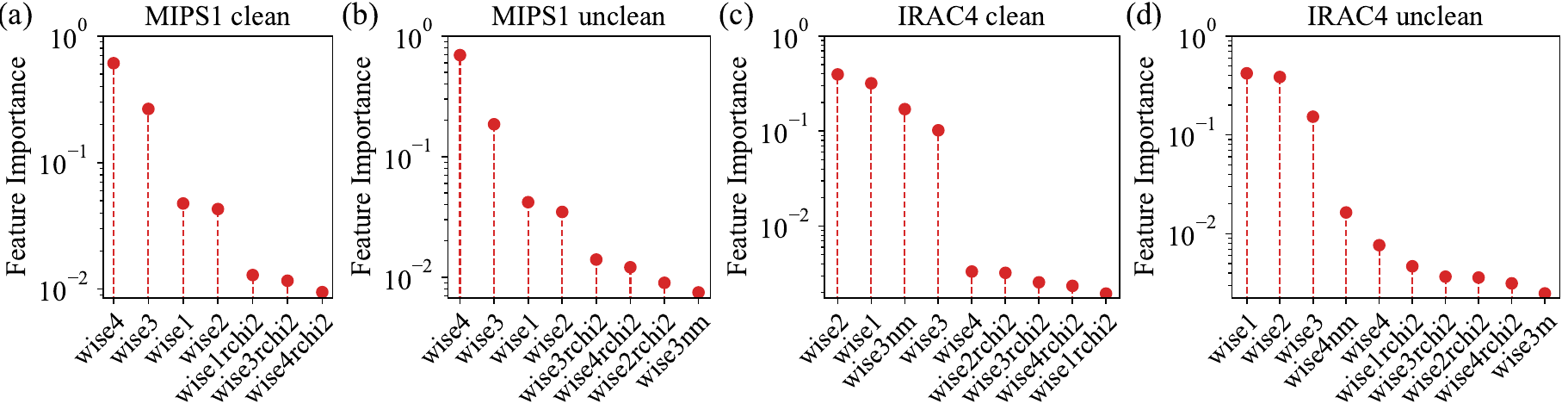}
\caption{WISE features selected by the Boruta algorithm and their importance in the ET models trained for the different studies. a) MIPS1 clean, b) MIPS1 unclean, c) IRAC4 clean, d) IRAC4 unclean.}
\label{fig:feature_importance}
\end{figure*}

As we stated in Section~\ref{sec:ML_features}, one of the most relevant aspects in ML modeling is to assess the importance of the different variables (features) involved in the prediction process. 
The Boruta algorithm found that the features relevant for the mid-infrared fluxes prediction task were the WISE fluxes at the four available channels (wise\#) and a variable number of quality flags (Fig.~\ref{fig:feature_importance}).
For the MIPS1 studies, this algorithm selected the reduced $\chi^2$ for the WISE bands 1, 3, and 4 (wise\#rchi2), while the frame coverage quality flags (wise\#m and wise\#nm) were deemed irrelevant (save for wise3nm in the unclean case).
For the IRAC4 studies, in contrast, some of the latter, together with wise2rchi2, were included into the selection of relevant features.

While the Boruta algorithm informs about the features that should be used for prediction, it is the ML regression model the one quantifying the predicting capability (i.e., importance) of each selected feature.
Figure~\ref{fig:feature_importance} shows bar plots with the importance of the features selected by the Boruta algorithm as calculated by the different ET models in this manuscript (Section~\ref{sec:ML_feature_impor}).
The feature selection step suggested that between seven and ten features --dependent on the study-- were relevant but, in spite of that, only three or four features seem to stand out from the rest. 
In general, the quality flags (wise\#rchi2 and wise\#m) are among the least predictive features (low importance).
In contrast, the flux densities are the ones that have more predictive weight (high importance).
In particular, for the MIPS1 studies, clearly the single most important feature is wise4, followed by wise3 and, even further with similar importances, wise1 and wise2.
One possible explanation to the relevance of wise4 in relation with MIPS1 is that, while they show wild discrepancies at low fluxes, they have almost an identical tendency when fluxes are high (Figs.~\ref{fig:ML_predictions}a and b). In a sense, the ET model ``knows'' that for high values of wise4, there is an almost one-to-one relation between W4 and MIPS1.
In contrast, for the lowest values this relationship does not hold, and the ET model must resort to other features to make the correct predictions, most likely wise3, and wise1 or wise2, followed by the quality flags to make fine adjustments.
Unfortunately, given the complexity of the ET model (100 different decision trees with dissimilar depths are used to construct it), we have not been able to find the exact value for this change in tendency.
Not that it is really necessary to find this value, as we will be systematically using the predicted fluxes whenever available in the cases that there is no \textit{Spitzer} observation.

As for IRAC4, wise1 and wise2 are the most predictive features, with very similar importance values, followed by wise3nm in the clean sample, and wise3 in the unclean sample (Figs.~\ref{fig:feature_importance}c and d). 
Given the trend observed for MIPS1, one could expect wise3, and not wise1, to be the most relevant feature for the IRAC4 study.
But this apparent contradiction makes sense, as IRAC4 and wise1, and for that matter wise2, show an almost perfect linear correlation for this dataset, with a Spearman correlation coefficient of $\rho\sim0.97$.
In contrast, the correlation between W3 and IRAC4 is more involved (Figs.~\ref{fig:ML_predictions}c and d). 
Thus, the corresponding ET model mostly relies on wise1, or wise2, to predict the fluxes at the IRAC4 band, and uses the remaining features (specially wise3nm or wise3), to refine the predictions.

Interestingly, wise1 and wise2 display almost the same importances across studies (both for MIPS1 and IRAC4), and this attends to their high correlation ($\rho\sim0.99$). 
Under these conditions, the ET model can use, almost interchangeably, wise1 and wise2 to make predictions, and, thus, it assigns the same importance to both. 
Nevertheless, it can happen that the ET models capture subtle differences between them, and leverage these subtleties to refine the predictions.

%
%------------------Summary and Conclusions----------------------

\section{\label{sec:conclusions}Summary and conclusions}

In this study, we present a novel method to make the most of WISE and \textit{Spitzer} capabilities taking full advantage of ML techniques. 
This approach has been proven to be very useful to find hidden relationships that exist between datasets, as the ones we analyze here. Specifically, we are using an ML regression model to predict mid-infrared fluxes at MIPS1 (24 $\mu$m) and IRAC4 (8 $\mu$m) bands from WISE variables (fluxes and quality flags).

Starting from a large sample of {\em Gaia} DR3  members of open clusters with WISE and \textit{Spitzer} data from SEIP\footnote{Updated versions of those samples and catalogs, for instance including {\em Gaia} DR4 members or enhanced WISE and \textit{Spitzer} data, could be used in the future if available.}, we first make use of a feature selection algorithm to choose the most important WISE features to improve the prediction quality. 
Then, after an exhaustive search of the finest ML algorithm for our predictive purposes, we focus in ET, as it gives the best performance (according to the values obtained for their goodness of fit metrics) but also provides us with an estimation of the prediction uncertainties (which is essential in astronomy).
We also use LR as a baseline model for comparison. 
We find that the goodness of fit metrics of the ET models are significantly better and the prediction uncertainties are rather small.

Throughout our study, we compare the results obtained for those members in SEIP with W4 nonzero (unclean sample) and those point sources among them with no contamination flags (clean sample). 
Our results show that data with better quality  give a better performance, which may seem obvious, although better quality means also less data for training. 
On the other hand, less quality contributes with irrelevant information. 
The best option should be a compromise between having enough quality (even though it may imply discarding some of the data) and keeping enough data for the best ML performance.
Therefore, we will continue using the predicted fluxes obtained from the clean data on our future analysis. 
We have confirmed the better concordance of the clean predicted fluxes in comparison with WISE and \textit{Spitzer} real ones by inspecting SEDs, corresponding to stars in the open cluster IC 348, built with a collection of predicted and observed fluxes.
We also ascertain that if, instead of predicting \textit{Spitzer} fluxes from selected WISE data using the ML approach described in this paper, we just carry a thorough selection of WISE fluxes following even more restrictive quality criteria (in order to avoid the discrepancies at lower fluxes), it results in an enormous loss of data and is therefore not recommended. 
In view of the above, we can conclude that our approach uses the best possible balance between data quality and quantity and, as a consequence, emerges as a riveting solution for other astrophysical studies when dealing with data discrepancies similar to the ones shown in this paper.\footnote{When applying this method to other samples or astrophysical studies, much care must be taken with their specific characteristics.}

\begin{acknowledgements}
      We thank the anonymous referee for the thorough reading of the manuscript and the helpful comments that improved the clarity and quality of this work.
      We warmly thank  M.C. Galvez-Ortiz for her fruitful discussion and initial assessment of the manuscript, and B. Montesinos for the numerous discussions helping to guide this work.
      L.C.~acknowledges partial support from a Ramón y Cajal Research Fellowship, Grant RYC2022-038362-I, funded by MCIN/AEI/10.13039/501100011033 and by the European Social Fund Plus (ESF+), and from Grant PID2020-114755GB-C31, funded by MCIN/AEI/10.13039/501100011033.
      This research has made use of the following resources: 
      The NASA/IPAC Infrared Science Archive, which is funded by the National Aeronautics and Space Administration and operated by the California Institute of Technology. 
      NASA's Astrophysics Data System Bibliographic Services. 
      The VizieR catalogue access tool, CDS, Strasbourg, France (DOI : 10.26093/cds/vizier). The original description of the VizieR service was published in \citet{Ochsenbein2000}.
      VOSA, developed under the Spanish Virtual Observatory (https://svo.cab.inta-csic.es) project funded by MCIN/AEI/10.13039/501100011033 through grant PID2020-112949GB-I00.
      VOSA has been partially updated by using funding from the European Union's Horizon 2020 Research and Innovation Programme, under Grant Agreement nº 776403 (EXOPLANETS-A).
      The steps and routines described here were implemented using Python's libraries \texttt{BorutaPy}, \texttt{NumPy} \citep{numpy}, \texttt{Pandas} \citep{pandas_paper, pandas_soft}, \texttt{PyCaret} \citep{PyCaret}, \texttt{scikit-learn} \citep{scikit-learn}, and \texttt{SciPy} \citep{scipy}.

\end{acknowledgements}

%
%------------------Bibliography-----------------

\bibliographystyle{aa} % style aa.bst
\bibliography{MLpapera_bib} % your references Yourfile.bib

%------------------End Bibliography-----------

%
%-------------------Appendix-------------------------

\begin{appendix}

\section{Data selection through WISE quality criteria}
\label{appendix_1}

\begin{figure*}[bhp]
\centering
\includegraphics[width=\linewidth]{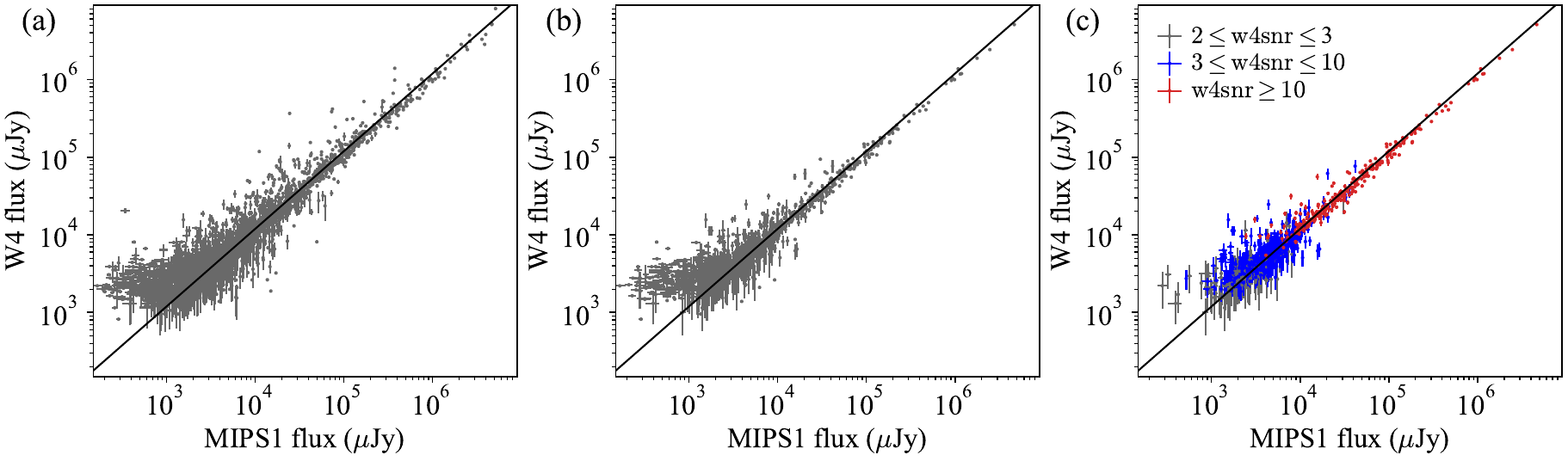}
\caption{Plots showing the three steps of the data selection procedure according to WISE quality criteria: a) W4 vs. MIPS1 unclean fluxes, b) W4 vs. MIPS1 clean fluxes, and c) W4 vs. MIPS1 fluxes with a S/N above 2 in the former. The different colors in c) correspond to three different S/N interval selections. Each flux is represented with its error bars. Diagonal black lines have been included in order to serve as reference. As in Figure \ref{fig:ML_predictions}, these lines are vertically shifted to take into account that both bands are measured at slightly different wavelengths.}
\label{fig:app}
\end{figure*}
\FloatBarrier

An interesting and thorough discussion about WISE contamination and source confusion can be found in \citet{Kennedy2012} and \citet{Dennihy2020}.
It becomes clear that an exhaustive ``cleaning'' of the samples following strict quality criteria, though necessary, would have come at a cost of an enormous and undesirable loss of data. 
With the aim to show this loss, and in order to compare the final amount of reliable data obtained under this method with those in our ML approach, we apply a selection procedure in sequential steps ({Fig.~\ref{fig:app}}). 
Our analysis here focuses on W4 data, as this band is the example of the worst-case scenario when filtering WISE data and also for showing the discrepancies with its \textit{Spitzer} equivalent, MIPS1. 
These discrepancies stand out when looking at the distances between the diagonal black lines and the points in every plot, that would be zero if both instruments were equally accurate. 
It is necessary to stress that a similar situation is found for W3 and IRAC4 ({Fig.~\ref{fig:ML_predictions}}), as well as in the remaining WISE bands and their \textit{Spitzer} equivalents, although not as overwhelming.

In order to proceed with our step-by-step selection procedure, we have no choice but to begin with the same objects of our unclean sample (Section~\ref{sec:sample}). This is, the 26068 sources in the SEIP Source List from the HR24 sample that have a nonzero value in the W4 band. 
A further selection of the sources with a nonzero value in MIPS1 (our reference data) leaves us with 2592 sources, less than the 10$\%$ of the initial number, a necessary restriction to be able to compare MIPS1 and W4 fluxes in our sample. 
The relation between these data and their notable discrepancies, mostly at the lowest values ($\textrm{MIPS1}\lesssim10^3\;\mu$Jy), can be seen in {Fig.~\ref{fig:app}a}.

The cleaning procedure that we are dealing with in this Appendix starts with the selection of point sources with no contamination flags (more details can be found in Section~\ref{sec:sample} and in \citealt{Kennedy2012}). 
This gives us the 1096 objects from our clean sample that have MIPS1 data. 
Remarkably, even after a massive loss in data, discrepancies between the fluxes are still observed, {Fig.~\ref{fig:app}b}.

It is noteworthy to mention that the unclean and clean samples represented here are the same ones we have been using as inputs for our ML models. This means that, in contrast to the procedure described in this Appendix, we could alternatively proceed with our ML approach with the advantage of not  causing more loss of data in our study. 
The plots here show that much care must be taken when using the direct value of the observed W4 fluxes in these samples. 
More stringent cut-offs in terms of the quality of the signal in the W4 band can be made by restricting the dataset to a signal-to-noise ratio w4snr$\geq$2, as seen in {Fig.~\ref{fig:app}c}. 
While this constraint improves the correlation by mainly removing the lowest fluxes, where the largest discrepancies are observed, it implies the loss of a large amount of data and, on top of that, the disparity does not entirely disappear.
For a better observation of the influence of the S/N in the fluxes correlation, we selected three different intervals, making a total of around just 650 sources: gray, between 2 and 3 (115 sources); blue, between 3 and 10 (347) and red for above 10 (286). 
For comparison, there are 508 sources with w4snr$\geq$3 and no detection in MIPS1 in the SEIP clean subsample (Fig.~\ref{fig:data_scheme}), well below the 4348 sources with predicted mid-infrared fluxes at MIPS1 24 $\mu$m band in the same subsample.
Thus, whereas increasing the S/N leads to better results, alternative methods such as our ML approach are needed to avoid losing almost all of the initial data (compromising the development of accurate studies with them) and, simultaneously, improve the correlations.

\end{appendix}

%-------------------End Appendix----------------------

\end{document}